\documentclass[useAMS,usenatbib]{mn2e}
\usepackage{graphicx}

\title[Scaling of the synchrotron cut-off]
{Scalings of the synchrotron cut-off and turbulent correlation of
active galactic nucleus jets}
\author[M.~Honda]{Mitsuru Honda$^{1}$\\
$^{1}$Plasma Astrophysics Laboratory, Institute for Global Science,
Mie, Japan}
\begin{document}

\date{Accepted 2010 July 15. Received 2010 June 24;
in original form 2010 February 4}

\pagerange{\pageref{firstpage}--\pageref{lastpage}} \pubyear{2010}

\maketitle

\label{firstpage}

\begin{abstract}
We propose a new analytic scaling of the cut-off frequency of
synchrotron radiation from active galactic nucleus (AGN)
jets that are nonuniformly filled with many filaments.
The theoretical upper limit is provided independent of magnetic
intensity, spectral index, coherence and correlation length of
filamentary turbulence, etc., such that
$\nu_{c}\simeq 6\times 10^{20}\delta[(r-1)/r]^{4/3}
(b/10^{-4})~{\rm Hz}$,
where $\delta$, $r$ and $b$ are the Doppler beaming factor,
shock-compression ratio and energy-density ratio of the
perturbed/local mean magnetic field of the filaments, respectively.
Combining our results with observational data for 18 extragalactic sources,
a constraint on the filament correlation length is
found, in order to give the number scaling of filaments.
The results suggest that, in particular, the jets of compact
BL\,Lacs possess a large number of filaments with
transverse size scale smaller than the emission-region size.
The novel concept of the quantization of
flowing plasma is suggested.
\end{abstract}


\section{INTRODUCTION}
\label{sec:1}

The pronounced extension of the synchrotron radio continuum to
the X-ray region, not formerly predicted, is now known to be a
very common feature among radio jets \citep[e.g.][]{harris06}.
The crucial point that should be emphasized is that, in many
sources, still no clear signs of sharp cut-off are found in the
{\it Chandra} regime (e.g. 3C\,66B: \citealt{hardcastle01};
M87: \citealt{marshall02}; 3C\,31: \citealt{hardcastle02};
M84: \citealt{harris02}; 3C\,346: \citealt{worrall05}),
and there is even an implication of the possible appearance of a
gamma-ray tail in the synchrotron spectral component.
This fact is of considerable interest, directly related
to the challenging issue of how high an energy range
electrons could be accelerated up to in situ.
In a conventional fashion, the electron energy distribution
is truncated at an assumed 'highest energy',
which could be responsible for the observed spectrum.
Independent of such a makeshift solution, however, it is
desirous to elaborate on a particle acceleration theory that can
naturally account for the extended continuum, with reference to
the morphological details innovated by the recent
very-long-baseline interferometry survey \citep[e.g.][]{lobanov01}.
From another point of view, theoretical expansion
would also be requisite for corroborating a unified scheme of
radio-loud active galactic nuclei \citep[AGNs;][]{urry95}.
None the less, the related thorough study required to elucidate
a universal mechanism underlying the many appearances of
the energetic continuum has not been performed as yet.

In a simplistic model assuming a homogeneous magnetic field
over a particular system, \citet{biermann87} first considered
a combination of small pitch-angle scattering diffusion and
Fermi-type acceleration of electrons to evaluate the frequency
of synchrotron emission from the highest energy electron:
\begin{equation}
\nu_{c}=3\times 10^{14}\left[3{\bar b}(U/c)^{2}\right]~{\rm Hz},
\label{eq:1}
\end{equation}
\noindent
where $U$, $c$ and ${\bar b}$ ($<1$) are the shock speed,
the speed of light and the ratio of the perturbed to the
mean energy density of the global magnetic field, respectively.
Although a symptom of cut-off in the range of equation~(\ref{eq:1})
has been suggested in some earlier publications
\citep[e.g.][]{keel88,perez-fournon88,meisenheimer96},
the broad-band spectral fitting is now found to, if anything,
entail a break followed by the aforementioned
high-energy extension, unlike a sharp cutoff.
A clue aiding in achievement of the required higher acceleration
efficiency might be the fine structure like filaments inside jets, which
can be reconciled with the observational detailed morphology.
Based on this notion, \citet{hh05a} have investigated
the root-mean-square (rms) diffusion of electrons in inhomogeneous
magnetic fields, and found that the intrinsic frequency
of the diffuse synchrotron component could reach as high as
\begin{equation}
\nu_{c}\sim 10^{24}B({\rm mG})[d({\rm pc})]^{2/3}
\left(U/c\right)^{4/3}~{\rm Hz},
\label{eq:2}
\end{equation}
\noindent
where $B$ and $d$ are the rms magnetic field strength and
transverse correlation length of filaments, respectively.
The relevant argument focusing on the specified case of
the nearby M87 jet has been given \citep{hh07},
although the value of $d$ remained unsolved.
Its determination is necessary not only to estimate
equation~(\ref{eq:2}) but also complete the filamentary jet model,
which could involve various radiation channels \citep{honda08}.

This paper has been prepared to spell out explicitly
the scalings of the cut-off and break for the dominant
synchrotron spectrum established via normal (non-diffuse)
processes in filamentary jets, and to revise equation~(\ref{eq:1}),
including relativistic beaming effects.
Making use of the results, we attempt to extract the scaling of
the lower limit of the filament correlation length, $d_{\rm min}$,
from observational data for sample extragalactic jets.
We find the property that the $d_{\rm min}$ value
increases as the size of emission regions increases.
A corollary derived from this is that the allowable maximum
number (capacity) of filaments with the outer scale ($\sim d$)
increases as the propagation distance of the jet increases.
For the situation in which AGN jets carry huge currents
driven by the central engines of black holes
\citep{appl92}, we take the current filamentation into
consideration in order to calculate the number of filaments ($N$).
From comparing $N$ with the capacity of the outer-scale filaments,
we infer the population of fine filaments inside the jets.
As a result, it is demonstrated that compact BL\,Lac
objects such as Mrk\,421 and 501 would possess typically
$\sim 10^{11}$ filaments with various transverse size
scales that are smaller than the entire emission-region size.

\section{THEORETICAL MODEL AND ANALYSIS}
\label{sec:2}

\subsection{The filamentary jet model}
\label{sec:2.1}

\begin{figure}
\centerline{\includegraphics*[bb=105.0 20.0 1035.0 760.0,
width=\columnwidth]{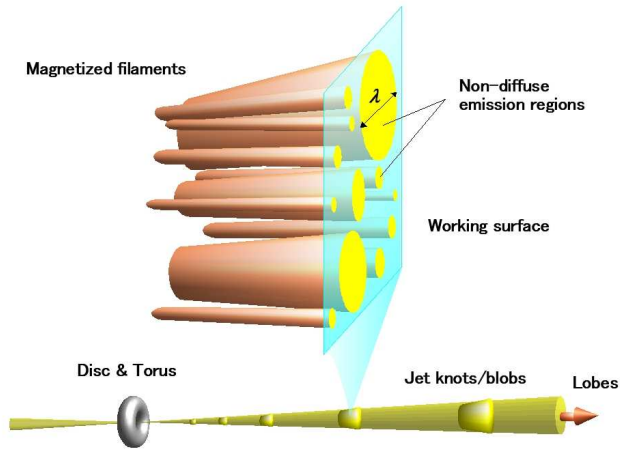}}
\caption{Schematic of the AGN jet that contains bright
knot-like features associated with shocks.
The magnification stands for a local sample around the
working surface accompanied by electron acceleration and
synchrotron emission.}
\label{fig:1}
\end{figure}

According to the original idea in \citet{hh04}, we rely on the
hypothesis that a jet consists of many magnetized filaments,
accommodated by some radio observations \citep[e.g.][]{owen89}.
For information, the circumstance concerned is
illustrated in Fig.~\ref{fig:1}.
The quasi-stationary magnetic fields are considered to be generated
by leptonic currents, and retain an energy-density level comparable
to that of the random component of lepton motion \citep{honda07}.
It is then expected that the Poynting flux level is less than
(or, at most, comparable to) the lepton energy flux.
In respect of the global energy budget, the small portion that
takes part in controlling the transverse dynamics
(Section~\ref{sec:5}) is loaded from the
energy reservoir of the jet bulk.
If the dominant kinetic energy is carried by hadronic components
then an excess of lepton energy departing from the mass ratio
(say, $0.05$ per cent for a simple electron--proton plasma
with no pairs) is entailed in order to generate the currents
and magnetic fields, in accord with arguments
concering the power of blazar jets \citep[e.g.][]{celotti08}.

The merging of current filaments can be associated with
the inverse cascade of turbulent magnetic energy.
For the inertial range of the turbulence, we invoke the
phenomenological expression for local magnetic intensity
$|{\bf B}|=B_{m}(\lambda/d)^{(\beta-1)/2}$, where $\lambda$
($\leq d$) is the transverse size of a filament (Fig.~\ref{fig:1})
and $\beta$ corresponds to the turbulent spectral index.
As seems plausible, this (zeroth order) magnetic field,
trapping lower-energy electrons, is disturbed by
Alfv\'enic waves to allow resonant scattering diffusion.
Also, in analogy to the knot and hotspot features, it is
reasonable to suppose a relativistic (non-relativistic)
shock overtaking the relativistic (non-relativistic) flow
with the Lorentz factor $\Gamma_{j}=(1-\beta_{j}^{2})^{-1/2}$,
such that the shock viewed in the upstream rest frame
likely has a non-/weakly relativistic speed.
Providing the fields defined upstream, we consider the
standard diffusive acceleration of bound electrons
due to the Fermi-I mechanism \citep{hh07},
which yields a power-law energy distribution
($\propto\gamma^{-p}$ for $\gamma\leq\gamma^{\ast}$).

\subsection{Updated scalings of break and cut-off frequencies}
\label{sec:2.2}

The acceleration time-scale $t_{\rm acc}$ is of the order of the
cycle time for one back-and-forth divided by the energy gain
per encounter with the shock \citep[e.g.][]{gaisser90}.
We here adopt a standard expression for $t_{\rm acc}$
that involves a negligible contribution from particle escape
downstream to the cycle time and gain \citep{ostrowski88,hh05b}.
What we deal with below is, therefore, the non-ageing
(or flaring) regime around the working surface (a related
issue is revisited in Section~\ref{sec:3.1}).
Compatible with this, it is reasonable to suppose that,
for major electrons bound to the outer-scale filament with
maximum field strength $\sim B_{m}$, the acceleration is degraded
by strong synchrotron cooling rather than escape losses \citep{honda08}.
The synchrotron emission, largely polarized, appears to constitute the
dominant radio continuum with a power index of $\alpha=(p-1)/2$
in the flux-density spectrum ($S_{\nu }\propto\nu^{-\alpha}$,
provided the electron density is uniform, e.g. \citealt{longair94}).
The spectral break reflects the highest energy of an
accelerated electron ($\gamma^{\ast}|_{\lambda\sim d}mc^{2}$)
at which $t_{\rm acc}$ is comparable to radiative time-scale.
From balancing these time-scales, we can analytically derive
the maximum Lorentz factor, $\gamma_{b}\equiv\gamma^{\ast}|_{\lambda=d}$
(along with the guidelines provided in \citealt{honda08});
for convenience, the generic expression is explicitly
written as follows:
\begin{equation}
\gamma_{b}=\left[f(b,r;B_{m},d)\right]^{1/(3-\beta^{\prime})}
\left[g(B_{m},d)\right]^{-(1+\beta^{\prime})/(3-\beta^{\prime})}.
\label{eq:3}
\end{equation}
\noindent
Here, $b$ is the energy density ratio of the perturbed/local mean magnetic
field of filaments, which is assumed to be a constant smaller than unity:
$b\ll 1$ (checked below), $r$ is the shock-compression ratio,
$\beta^{\prime}(>1)$ is the turbulent spectral index of the Alfv\'enic
fluctuations (superimposed on the local mean field) and the
dimensionless functions are defined as
\begin{equation}
f\equiv 9\pi^{2}(\beta^{\prime}-1)b\frac{r-1}{r}
\frac{B_{m}d^{2}}{e};~~~
g\equiv\frac{eB_{m}d}{2mc^{2}},
\label{eq:4}
\end{equation}
\noindent
where $c$, $e$ and $m$ are the speed of light, elementary
charge and electron rest mass, respectively.
The observed break frequency can be simply estimated as
$\nu_{b}=(3/4\pi)\delta\gamma_{b}^{2}[eB_{m}/(mc)]$,
where $\delta=\Gamma_{j}^{-1}(1-\beta_{j}\cos\theta)^{-1}$
is the beaming factor.
When setting the standard value of $\beta^{\prime}=\frac{5}{3}$
(for Kolmogorov turbulence), we obtain the scalings
\begin{equation}
\gamma_{b}=1.1\times 10^{6}\xi_{-5}^{3/4}B_{m}^{-5/4}d^{-1/2},
\label{eq:5}
\end{equation}
\begin{equation}
\nu_{b}=5.2\times 10^{15}\delta\xi_{-5}^{3/2}B_{m}^{-3/2}d^{-1}~{\rm Hz},
\label{eq:6}
\end{equation}
\noindent
where $\xi_{-5}=\xi/10^{-5}$ and $\xi=b[4\left(r-1\right)/3r]$,
$B_{m}$ in mG and $d$ in pc.
Note that both equations~(\ref{eq:5}) and (\ref{eq:6})
are independent of $\beta$.

In filaments smaller than $d$, the bound electrons
can be accelerated up to a higher energy, i.e.
$\gamma^{\ast}|_{\lambda<d}>\gamma_{b}$,
owing to the weaker synchrotron loss.
While the synchrotron flux density is lower, the power-law
tail is retained up to higher frequency
(reflecting the enhanced $\gamma^{\ast}$ value).
Apparently, this property can be responsible for
the extended continuum going beyond $\nu_{b}$.
However, the increase in acceleration efficiency ought to be
limited at a critical coherent length $\lambda_{c}$, below which
the transverse escape of electrons dominates the radiative loss.
In the escape-dominant regime, another $\gamma^{\ast}$ scaling
is derived from the spatial limit condition, i.e. equating
the electron gyroradius $r_{g}$ with $a(\lambda/2)$, where
$a$ is a dimensionless factor smaller than unity
(see Section~\ref{sec:2.3}).
The equation for
$\gamma^{\ast}|_{\lambda<\lambda_{c}}=\gamma^{\ast}|_{\lambda>\lambda_{c}}$
contains the solution of $\lambda=\lambda_{c}$ at which
$\gamma^{\ast}$ takes a peak value, to give the expression
$\lambda_{c}/d=a^{2(3-\beta^{\prime})/(3\beta+1)}
f^{2/(3\beta+1)}g^{-8/(3\beta+1)}$.
Substituting the $\lambda_{c}$ expression into the $\gamma^{\ast}$
scaling, the critical Lorentz factor, defined as
$\gamma_{c}\equiv\gamma^{\ast}|_{\lambda=\lambda_{c}}$,
is obtained as
\begin{equation}
\gamma_{c}=a^{[(\beta+1)\beta^{\prime}-2]/(3\beta+1)}
f^{(\beta+1)/(3\beta+1)}g^{-(\beta+3)/(3\beta+1)},
\label{eq:7}
\end{equation}
\noindent
and, in turn, the intrinsic cut-off frequency
(in the observer frame) can be estimated as
$\nu_{c}=(3/4\pi)\delta\gamma_{c}^{2}[eB_{m}/(mc)]
(\lambda_{c}/d)^{(\beta-1)/2}$, which is found to
be recast in the simple form
\begin{equation}
\nu_{c}=27\pi(\beta^{\prime}-1)\delta
a^{\beta^{\prime}-1}b\frac{r-1}{r}\frac{mc^{3}}{e^{2}},
\label{eq:8}
\end{equation}
independent of $\beta$, $B_{m}$, $d$, etc.
For $\beta^{\prime}=5/3$, the quantities scale as
\begin{equation}
\gamma_{c}=3.8\times 10^{8}a^{3/7}\xi_{-5}^{3/7}B_{m}^{-2/7}d^{1/7}
\label{eq:9}
\end{equation}
\noindent
for $\beta=2$ (see \citealt{montgomery79} as an example) and
\begin{equation}
\nu_{c}=4.5\times 10^{19}a^{2/3}\delta\xi_{-5}~{\rm Hz}
\label{eq:10}
\end{equation}
\noindent
respectively.
Note that equation~(\ref{eq:10}) is independent of $\beta$.
In the ordinary case in which the diffuse emission level is
lower (cf. the discussion in \citealt{hh07}), equation~(\ref{eq:10})
(instead of equation~\ref{eq:2}) provides the
observed cut-off frequency of the synchrotron radiation,
which is of non-diffuse at least in the scale of $\sim\lambda_{c}$.
Note that the absorption in extragalactic background lights is
expected to be insignificant, as long as the radiation frequency is
in the range below $\sim 10^{24}~{\rm Hz}$ \citep[e.g.][]{kneiske04}.

\subsection{The smearing effect on the spectrum at $\nu_{c}$}
\label{sec:2.3}

The large $\nu_{c}$ value expected in equation~(\ref{eq:10})
is owed to the efficient acceleration of electrons bound to
the local magnetic fields of filaments with
size-scale $\sim\lambda_{c}$.
In this regime, the time-scale of escape from a small-scale
filament competes with the radiative loss time-scale, and therefore
we provide additional discussions concerning the spatiotemporal
property of diffusion loss, in order to identify the value of $a$.
The escape is dominated by diffusion across the magnetic
field line permeating through the current filaments, so that
for the electrons confined within radial size
$\sim\lambda/2$, the escape time is estimated as
$t_{\rm esc}\sim(\lambda/2)^{2}/c\ell_{\perp}$, where $\ell_{\perp}$
is the mean free path (mfp) for diffusion perpendicular
to the magnetic field line \citep[e.g.][]{hillas84}.
Recall that the perpendicular diffusion is
characterized by the coefficient
$\kappa_{\perp}=\kappa_{\parallel}/(1+\eta^{2})$,
where $\eta=\ell_{\parallel}/r_{g}$,
$\kappa_{\parallel}=\frac{1}{3}\eta r_{g}c$ and
$\ell_{\parallel}=r_{g}/[b(\beta^{\prime}-1)]
[\lambda/(2r_{g})]^{\beta^{\prime}-1}$ (for $\beta^{\prime}>1$)
denotes the mfp for the diffusion parallel to
the field line \citep[e.g.][]{muecke01,hh05b}.
Along with this, we invoke the relation
$\ell_{\perp}\simeq\eta^{-2}\ell_{\parallel}$
for $\eta^{2}\gg 1$, which is the region of interest
to us, and then the escape time can be expressed as
$t_{\rm esc}=(\ell_{\parallel}/c)(\lambda/2r_{g})^{2}$.
As a result, we find that equating $t_{\rm esc}$ to
$t_{\rm acc}$ correctly provides the aforementioned spatial
limit equation in the form of $r_{g}=a(\lambda/2)$, where
\begin{equation}
a=\sqrt{\left(r-1\right)/r}.
\label{eq:11}
\end{equation}
One can ascertain that equation~(\ref{eq:11}) is
also valid for $\beta^{\prime}=1$ (not shown).
In the strong-shock limit we have $a=0.87$, and suggest
that setting $a$ to unity is nearly adequate for
the simple treatment of escape loss.
In the case in which the mfp related to the diffusion
considered is anomalously longer than the estimated range
of $\ell_{\perp}$, the $a$ value becomes effectively
smaller than that of equation~(\ref{eq:11}); i.e.
$\gamma_{c}$ and $\nu_{c}$ could reduce, even if the inverse
Compton (IC) and redshift effects considered later are negligible.
The uncertainty in $a$ is expected to cooperate to
smear out the actual spectrum around $\nu_{c}$.

\subsection{The inverse Comptonization effect}
\label{sec:2.4}

There exists another possible effect that prevents electrons from
being energized up to the range given in equation~(\ref{eq:9}).
The starlight and dust emission that emanate from
host galaxies potentially serve as external targets
for the inverse Comptonization of accelerated electrons.
For relativistic jets, one may consider these and also the
cosmic microwave background, because the radiation
energy density is boosted in the comoving frame by
a factor of $\Gamma_{j}^{2}$ \citep{stawarz03}.
Particularly in the weaker magnetic intensity regions of
$\lambda\ll d$, the comoving radiation energy density
(denoted as $u_{\rm rad}$) could be comparable to the
magnetic energy density, so that the time-scale for Thomson
scattering loss could compete with that for synchrotron loss.
It is pointed out that similar circumstances can appear
even for a synchrotron originator, if the non-local effect
is significant for radiative transfer.
On the other hand, it is unlikely that inverse
Comptonization affects the lepton energization to
the $\gamma_{c}$ level of equation~(\ref{eq:9}),
because the Klein-Nishina effect tends to reduce
the scattering cross-section \citep[e.g.,][]{hh07}.
In any case, the competition between these conceivable radiative channels
is complicated, dependent on the detailed parameters inherent to sources.

With this aspect in mind, we here simply clarify the maximum level of
IC losses that can affect the electron acceleration up to $\gamma_{c}$.
For example, one can identify the maximum equivalent IC field
(denoted as $B_{\rm ic}$) corresponding to the quantity
$\sqrt{8\pi u_{\rm rad}}$ by conservatively
taking the balance of $t_{\rm acc}$ with the
Thomson time-scale at $\gamma_{c}$.
Considering the synchrotron limit at $\gamma_{b}$,
it is found that any IC losses can safely be ignored
(i.e. equations~\ref{eq:9} and \ref{eq:10} are valid) if
\begin{equation}
\frac{B_{\rm ic}}{B_{m}}<\left(\frac{\gamma_{b}}{\gamma_{c}}\right)^{
(\beta-1)(3-\beta^{\prime})/[(\beta+1)\beta^{\prime}-2]}.
\label{eq:12}
\end{equation}
The criterion for $(\beta,\beta^{\prime})=(2,5/3)$ reduces to
$B_{\rm ic}/B_{m}<(\gamma_{b}/\gamma_{c})^{4/9}$,
which would generally be satisfied.
In the following, therefore, we are concerned with the negligible
contribution of Compton losses to the maximum energy analysis,
aiming at providing a generic theoretical upper limit on the
(non-diffuse) synchrotron radiation frequency
that has not been manifested so far.

\section{COMPARISON WITH OBSERVATIONAL DATA}
\label{sec:3}

\subsection{The ageing effect and degenerative $\nu_{c}$ redshift}
\label{sec:3.1}

\begin{figure}
\centerline{\includegraphics*[bb=110.0 80.0 708.0 530.0,
width=\columnwidth]{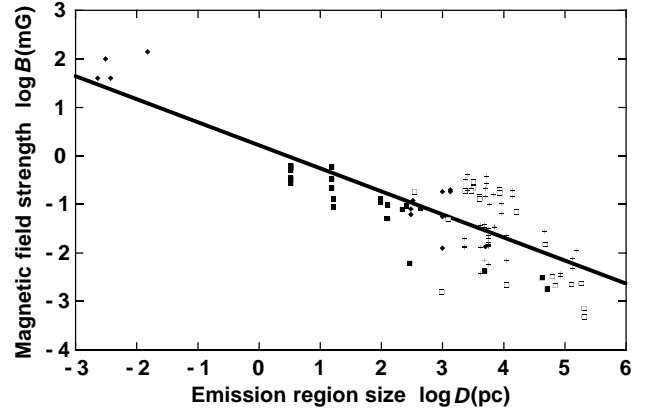}}
\caption{Magnetic field strength $B$, versus emission-region size $D$,
for blazars (filled diamonds), FR\,I (filled squares) and
FR\,II (open squares) radio sources and QSOs (crosses).
$B$ equipartition values in the rest frame,
i.e. $B_{{\rm eq},\delta=1}\delta^{-5/7}$, are from
\citet{kataoka05} (KS05), except for the marginal data of four
compact BL\,Lacs, which are commonly from the synchrotron self-Compton
models provided in the literatures in Table~\ref{tbl:1}.
It is noted that for the 'SYN' features classified in KS05,
$B$ is simply set to $B_{{\rm eq},\delta=1}$.
The solid line indicates the best-fitting scaling
$B=1.7D^{-0.48}$.}
\label{fig:2}
\end{figure}

We further expand our discussions on the validity of the
scalings of $\nu_{b}$ and $\nu_{c}$,
in light of the comparison with observational data.
In equation~(\ref{eq:6}), we have the $\nu_{b}$ scaling
apparently proportional to $B_{m}^{-3/2}$.
However, there is a potential that the correlation length-scale
$d$ is virtually related to $B_{m}$, yielding an
explicit magnetic intensity dependence of $\nu_{b}$.
To see this, in Fig.~\ref{fig:2}, we plot magnetic field
strength $B$ against emission-region size $D$ for
48 sample objects (102 features) that include
blazars, Fanaroff-Riley type I/II radio galaxies (denoted as
FR\,I/\,II, respectively) and quasi-stellar objects (QSOs).
It is found that the power-law fitting suggests a generic
scaling of $D\sim B^{-2.1\sim(-2)}$, though the
translation to a $d-B_{m}$ relation is not trivial.
At this juncture, provided the scaling is the same, one may
read off $\nu_{b}\sim B_{m}^{0.5\sim 0.6}$; when supposing a
simple ordering of $B/B_{m}\la 1$ and that $d\sim D^{0.7}$
is retained (cf. Section~\ref{sec:4}), we preliminarily
obtain $\nu_{b}\sim B_{m}^{-0.1\sim +0.1}$.
In any case, the resulting magnetic intensity dependence of
$\nu_{b}$ is thought of as being weak in the present context.
This property appears to be, if it is correct, markedly different
from $\nu_{b}\propto B^{-3}$ \citep{brunetti03,cheung05},
which was derived considering the synchrotron cooling
of ageing electrons in hot spots \citep{meisenheimer97}.
We point out that observations provide some support for
the classical scaling (for hotspots), implying that the
present analytic results cannot directly be applied
to hotspot phenomena, although the applicability of
the filament model as such still remains unclear.

In the synchrotron spectrum, the appearance of ageing effects
might be pronounced, particularly during the low activity around
the working surface, which can be translated as the quenching
of turbulent energy injection at the smallest size-scale
\citep[i.e. the largest wavenumber, $k_{\rm m}$,
in the filamentary turbulent spectrum:][]{honda00b}.
The inertial inverse cascade will lower the upper cut-off
$k_{\rm m}$, resulting in a lack of smaller-scale filaments
and magnetic energy condensation around the outer scale $\sim d$.
When entering into the degenerate regime of
$k_{\rm m}^{-1}>\lambda_{c}/2\pi$,
equation~(\ref{eq:10}) is violated, and the further cascade leads to
the 'redshift' of cut-off frequency moving toward the lower limit
compared with the $\nu_{b}$ range given in equation~(\ref{eq:6}).
This scenario provides a reasonable interpretation of the observational
fact that there are still many objects with a cut-off signature
around the optical band \citep[e.g.][]{meisenheimer97,mack09}, while
there is evidence of synchrotron X-ray emission in a number
of jets and hotspots \citep{harris06}; i.e. the
inspected low-energy cut-offs do not always imply
extremely small values of $\xi$ in equation~(\ref{eq:10}).
In the specific regime, a spectral break below the
low-energy band could reflect the oldest electron population
along the classical scenario, whereupon the scaling of
$\nu_{b}\propto B^{-3}$ might be realized
\citep{brunetti03,cheung05}.

\subsection{Theoretical upper limit of $\nu_{c}$}
\label{sec:3.2}

Listed in Table~\ref{tbl:1} are the relevant data for radio sources,
which are used to evaluate the value of $\nu_{c}$ in equation~(\ref{eq:10})
and to compare the measured $\nu_{b}$ with equation~(\ref{eq:6}).
Basically, we have sought X-ray sources that retain the
extension of the synchrotron continuum above the confirmed
$\nu_{b}$, although for the data selection the
statistical bias might inevitably be involved.
The samples selected here include BL~Lac objects,
FR\,I radio galaxies, core-dominated quasars (CDQ) and a
gamma-ray quasar (GRQ); these are found, in particular, to provide
a well-defined parameter set that could also be referred to the
correlation length estimate later in equation~(\ref{eq:13}).
Taking a conservative approach, FR\,II sources have been excluded at the moment,
since hotspot phenomena in the tips often incur the ageing effects that
have been annotated in Section~\ref{sec:3.1}, and also show some
peculiar features in energetic emission \citep[e.g. Pic\,A;][]{tingay08}.
On the other hand, there are some cases for which FR\,II features
resemble knots in jet \citep[e.g.][]{meisenheimer97,marshall10},
and therefore the applicability of the present model to
FR\,II sources will be investigated in more detail,
though it seems to be somewhat beyond the scope of this paper.

\begin{table*}
 \centering
 \begin{minipage}{165mm}
  \caption{Sample parameters to evaluate equations~(\ref{eq:10})
           and (\ref{eq:13}).\label{tbl:1}}
  \begin{tabular}{@{}lccccccc@{}}
  \hline
   Host/feature\,(class) &
   $D$(pc)\footnote{Values set simply to $D=L/50$, if not well defined.} &
   $p$\footnote{Values inferred from comparing the measured radio
                spectral indices with $\alpha=(p-1)/2$.} &
   $B_{m}$(mG)\footnote{For an ad hoc manner, set to ten times
                        the field strength suggested in Refs
                        \citep[for the reasoning, cf.][]{honda08}.} &
   $\delta$  & $\nu_{b}$(Hz)  & $\nu_{c}$(Hz) &
   References\footnote{Ref: 1. \citealt{krawczynski04},
2. \citealt{konopelko03}, 3. \citealt{blazejowski05},
4. \citealt{tavecchio01}, 5. \citealt{urry97},
6. \citealt{kataoka00}, 7. \citealt{zhang02},
8. \citealt{dominici04}, 9. \citealt{pesce01},
10. \citealt{kataoka05}, 11. \citealt{harris06},
12. \citealt{harris02}, 13. \citealt{hardcastle02},
14. \citealt{tansley00}, 15. \citealt{hardcastle01},
16. \citealt{owen89}, 17. \citealt{biretta91},
18. \citealt{perlman01}, 19. \citealt{marshall02},
20. \citealt{waters05}, 21. \citealt{kataoka06},
22. \citealt{hardcastle06}, 23. \citealt{worrall05},
24. \citealt{whiting03}, 25. \citealt{schwartz06},
26. \citealt{sambruna01}, 27. \citealt{jester02},
28. \citealt{sambruna04}, 29. \citealt{schwartz00},
30. \citealt{chartas00}.}\\
 \hline
1ES\,1959+650\,(BL\,Lac) & $3.8\times 10^{-3}$ & $2$ & $400$ &
$20$ & $10^{16}-10^{17}$ & $\geq 10^{19}$ & 1\\
Mrk\,421\,(BL\,Lac) & $3.1\times 10^{-3}$ & $1.75$ & $1000$ &
$55$ & $10^{16}$ & $>10^{19}$ & 2,3\\
Mrk\,501\,(BL\,Lac) & $2.3\times 10^{-3}$ & $1.60$ & $400$ &
$50$ & $10^{16}$ & $\geq 10^{20}$ & 2,4\\
PKS\,2155-304\,(BL\,Lac) & $1.5\times 10^{-2}$ & $1.35$ & $1400$ &
$28$ & $(.5-1.3)\times 10^{16}$ & $>10^{19}$ & 5,6,7,8\\
3C\,371/A\,(BL\,Lac) & $252$ & $2.52$ & $.81$ &
$6$ & $10^{14}-10^{15}$ & $>10^{17}$ & 9,10,11\\
M84/N3.3\,(FR\,I) & $8$ & $2.3$ & $1.3$ & $.6-1.25$ &
$4\times 10^{13}$ & $>2\times 10^{17}$ & 10,11,12\\
3C\,31\,(FR\,I) & $68$ & $2.1$ & $1.1$ &
$1.3$ & $10^{13}$ & $\geq 2\times 10^{17}$ & 10,11,13\\
3C\,66B/B\,(FR\,I) & $85$ & $2.2$ & $1.2$ &
$1.2-2.8$ & $10^{14}$ & $>2\times 10^{17}$ & 11,14,15\\
M87/A\,(FR\,I) & $110$ & $2.34$ & $5.1$ &
$1-3$ & $(.27-7.0)\times 10^{16}$ & $>2\times 10^{17}$ & 16,17,18,19,20\\
Cen\,A\,(FR\,I) & $154$ & $2$ & $6.4$ &
$4$ & $10^{12}-10^{14}$ & $\geq 2\times 10^{17}$ & 10,11,21,22\\
3C\,346\,(FR\,I) & $320$ & $2$ & $7.2$ &
$<3$ & $10^{12}-10^{13}$ & $\geq 2\times 10^{17}$ & 11,23\\
PKS\,0208-512/R2\,(GRQ) & $5240$ & $2.4$ & $.135$ & $7.5$ &
$\leq 10^{14}$ & $\geq 10^{17}$ & 11,24,25\\
3C\,273/D\,(CDQ) & $3000$ & $2.6$ & $220$ &
$4.2$ & $10^{13}-10^{14}$ & $>10^{15}$ & 26,27\\
4C\,49.22/B\,(CDQ) & $7800$ & $2.4$ & $.75$ &
$14$ & $10^{13}-10^{14}$ & $>10^{15}$ & 10,11,28\\
4C\,19.44/A\,(CDQ) & $10000$ & $2.4$ & $.86$ &
$14$ & $10^{13}$ & $>10^{15}$ & 10,11,28\\
PKS\,1202-262/R2\,(CDQ) & $11400$ & $2.4$ & $120$ &
$12$ & $10^{11}-10^{14}$ & $\geq 10^{15}$ & 11,25\\
PKS\,1030-357/R2\,(CDQ) & $13600$ & $2.4$ & $223$ &
$9.2$ & $10^{11}-10^{14}$ & $\geq 10^{15}$ & 11,25\\
PKS\,0637-752\,(CDQ) & $16700$ & $2.62$ & $2$ &
$10$ & $<3\times 10^{12}$ & $\geq 5\times 10^{14}$ & 11,29,30\\
\hline
\end{tabular}
\end{minipage}
\end{table*}

The updated X-ray data provide the lower limits on $\nu_{c}$
in equation~(\ref{eq:10}), and thereby the lower limits
of the unfixed parameter, $\xi_{\rm min}$,
for given $a\sim 1$ (upper limit) and measured $\delta$ values.
Then, by invoking a simple relation $p\sim(r+2)/(r-1)$
\citep[see][for details]{schlickeiser02},
the lower limits of $b$ can be estimated such that
$b_{\rm min}\sim[(p+2)/4]\xi_{\rm min}$.

\begin{figure}
\centerline{\includegraphics*[bb=146.0 20.0 645.0 545.0,
width=\columnwidth]{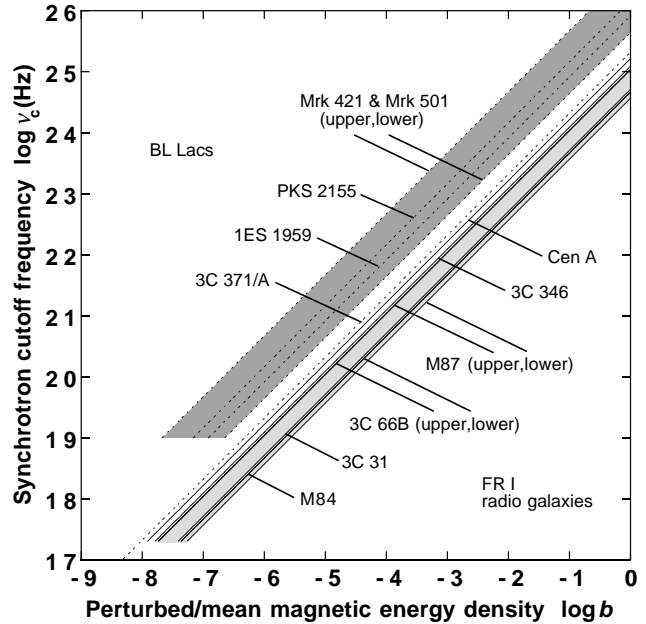}}
\caption{The cut-off frequency $\nu_{c}$ of synchrotron
emission from the extragalactic jets (labelled)
as a function of the energy density ratio of the
perturbed/local mean magnetic field of filaments, $b$.
The darkly and lightly shaded areas indicate the expected
domains for Mrk\,421 and 501 with $(p,\delta)=(1.6-1.8,10-100)$
and M87 with $(2.3,1-3)$, respectively.}
\label{fig:3}
\end{figure}

Fig.~\ref{fig:3} plots the predicted values of $\nu_{c}$
against $b$ (in the range of $\geq b_{\rm min}$),
except for CDQs (for reasoning, see Section~\ref{sec:4}).
According to equation~(\ref{eq:11}), the value of $a$ is
simply set to $\sqrt{3/(p+2)}$ hereafter.
It is confirmed that the condition of $b\ll 1$
(introduced in Section~\ref{sec:2.2})
is satisfied as long as $\nu_{c}\leq 10^{23}~{\rm Hz}$.
For a given $b$, the $\nu_{c}$ values of the compact BL\,Lacs
are likely higher than those of the FR\,I sources, mainly
because of the higher $\delta$ values for the BL\,Lacs.
For the set $(p,\delta)=(1.6,10)$ \citep[Mrk\,421,
Mrk\,501: e.g.][]{tavecchio01,konopelko03,blazejowski05},
a specified value of $b=1\times 10^{-3}$ yields
$\nu_{c}=4.7\times 10^{22}~{\rm Hz}$, in conformity with
the extremely high-frequency ranges of synchrotron cut-off
expected in TeV gamma-ray emitters \citep{honda08}.
It is also mentioned that the scaling of a beaming GRQ
\citep[PKS\,0208-512 with a hundred-kpc scale;][]{schwartz06}
appears to be close to that of a kpc-scale BL\,Lac
\citep[3C\,371:][not shown in the figure]{pesce01}.

\section{CORRELATION LENGTH-SCALE AND 'PACKAGE' OF FILAMENTS}
\label{sec:4}

A worthwhile manipulation is to combine equation~(\ref{eq:6})
with (\ref{eq:10}), eliminating the parameter $\xi$.
We solve this for the correlation length $d$
to find the scaling
\begin{equation}
d=5.4\times 10^{-3}a^{-1}
\delta^{-1/2}\nu_{b,14}^{-1}\nu_{c,17}^{3/2}B_{m}^{-3/2}~{\rm pc},
\label{eq:13}
\end{equation}
where $\nu_{b,14}=\nu_{b}/10^{14}{\rm Hz}$ and
$\nu_{c,17}=\nu_{c}/10^{17}{\rm Hz}$.
By putting the parameter values of Table~\ref{tbl:1}
into the right-hand side of equation~(\ref{eq:13}),
one can fix the allowed $d$ domain for each sources.
In the derivation of equation~(\ref{eq:13}), it is supposed that
the parameter $\xi$ in equations~(\ref{eq:6}) and (\ref{eq:10})
takes a common value \citep[cf.][]{honda08}.
If the back-reaction effect of accelerated particles on the
shock structure \citep[e.g.][]{blasi02,kang07} comes into
play with a positional dependence, the value of $\xi$ may be
different in between the regions of $\lambda\sim d$ and
$\lambda_{c}$, as $b$ and $r$ change spatially.
Related to this issue, the strong non-linear effect also
invalidates the assumption of $b\ll 1$ followed by $\eta\gg 1$.
We mention here that the back-reaction effect may scatter
the results shown below, resulting in an effective
lower estimate of $d$ (in equation~\ref{eq:13}), but
quasi-linearity ($b\ll 1$) seems to be almost
satisfied even in the smaller filament with $\lambda_{c}$,
at which the acceleration efficiency is maximum.

\begin{figure}
\centerline{\includegraphics*[bb=42.0 50.0 558.0 730.0,
width=\columnwidth]{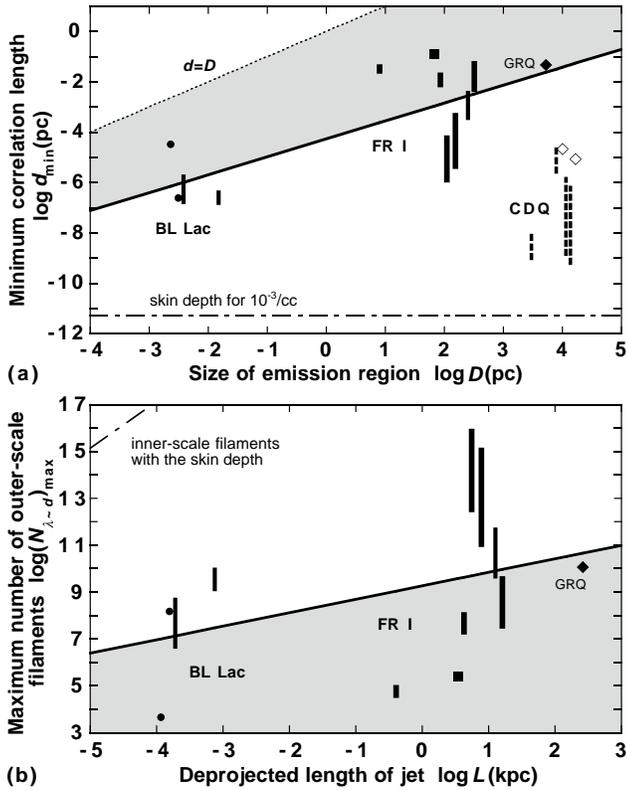}}
\caption{(a) The lower limit of the transverse
correlation length of filamentary turbulence
$d_{\rm min}$ versus the emission-region size $D$.
The shaded area between the spatial limit (dotted line)
and a power-law scaling (solid line) indicates the
allowed domain for the correlation length.
A skin depth is also indicated (e.g. for a plasma density
of $10^{-3}~{\rm cm^{-3}}$; dot--dashed line).
(b) The maximum number of outer-scale filaments
$(N_{\lambda\sim d})_{\rm max}$ versus the
deprojected length of the jet $L$.
The shaded area (below the scaling; solid line)
suggests the capacity of the outer-scale filaments.
The dot-dashed line indicates an upper limit: $D^{2}$ divided
by the square of the skin depth (for $10^{-3}~{\rm cm^{-3}}$).
BL~Lac: circles, bars; FR\,I: filled square, fat bars;
GRQ: filled diamond; CDQ: open diamonds, dotted bars.}
\label{fig:4}
\end{figure}

Fig.~\ref{fig:4}(a) plots the lower limit, $d_{\rm min}$,
against the size of emission region $D$.
Note the relation $d_{\rm min}<D$, which maintains
the theoretical consistency.
As seen in the figure, the plotted points including
bars may be separated roughly into three groups
(labelled as BL\,Lac, FR\,I and CDQ).
Notice that the FR\,I group apparently absorbs a large-scale
BL\,Lac (3C\,371), implying the similarity of turbulent states.
Although the uncertainty ascribed to the observations is
not small for the moment, it seems natural to
suppose that the $(D,d_{\rm min})$ points of the BL~Lac
and FR\,I share a common power-law scaling.
The CDQ group is apparently isolated in a marginal region below the
scaling (mainly because of the larger uncertainty of $\nu_{c}$,
stemming from the flat spectral features; cf. Table~\ref{tbl:1}),
although the $d_{\rm min}$ levels are still retained above a skin depth
for the plasma density of $>10^{-6}~{\rm cm^{-3}}$, satisfying
the restriction $\lambda/d_{\rm min}\leq 1$ \citep[consistent
with the filamentation instability;][]{honda00b}.
With the exception of the marginal data,
the power-law fitting yields a scaling of
$d_{\rm min,pc}\sim 5\times 10^{-5}D_{\rm pc}^{0.7}$,
where the units are pc.
This arguably reflects the history of how the filamentary
turbulence evolves as the jet propagates, increasing the width.
The physical implication can be revealed by translating
a quantity $\la D^{2}/d_{\rm min}^{2}$ as the
upper limit of the number of outer-scale filaments,
$(N_{\lambda\sim d})_{\rm max}$.

Fig.~\ref{fig:4}(b) plots the values of
$(N_{\lambda\sim d})_{\rm max}$ ($=D^{2}/d_{\rm min}^{2}$, for
convenience) as a function of the deprojected length of jets $L$.
Interpolating them yields the scaling
$(N_{\lambda\sim d})_{\rm max}\sim 10^{9}L_{\rm kpc}^{0.6}$,
indicating the trend that the capacity of outer-scale
filaments increases as $L$ increases (shaded area).
In order to see the significance, we recall a promising scenario
in which jets having large-scale magnetic fields are necessarily
accompanied by huge currents in the bulk plasmas \citep{appl92},
the kinetic energy of which is dissipated only a little during
the transport from the cores to the lobes \citep{tashiro04}.
As a whole, this is compatible with the superconductivity
of plasmas.

For example, let us consider the well-confirmed samples
M87 and Cen\,A, the nuclei of which have the supermassive black holes
with mass $\sim 10^{9}M_{\sun}$ \citep{macchetto97} and
$(10^{7}-10^{8})M_{\sun}$ \citep{marconi06}, respectively.
According to the arguments of \citet{appl92}, their central
engines, incorporated with their accretion discs, ought to have
the potential to drive a current of the order of magnitude of
$I\sim 10^{19}~{\rm A}$ and $\sim 10^{18}~{\rm A}$, respectively.
Such a huge current could not be transported by a single uniform
column, on account of the current inhibition \citep{honda07}.
One possible solution is to allow for the presence of many
filaments that each carry a current limited by
$i_{0}\sim (mc^{3}/e)\beta_{j}\Gamma_{j}$ \citep{honda00}.
It is noted that the value of $i_{0}$ is independent of
$\lambda$ \citep{honda00b}.
The number of current filaments can then be estimated as
$N(\sim I/i_{0})\sim 10^{15}$ and $\sim 10^{14}$, respectively.
Note that these values are just in the expected ranges of
$(N_{\lambda\sim d})_{\rm max}$ for M87 and Cen\,A
(cf. Fig.~\ref{fig:4}b).
The outcome suggests that, if the actual capacity of the outer-scale
filaments reaches the level $\sim (N_{\lambda\sim d})_{\rm max}$,
the filament cluster, which consists of the smaller filaments
with size $\lambda<d$, would not be closely packed in the jet.
This seems to be qualitatively amenable to the observed
appearance of the non-uniformly filled features in the
jets of M87 \citep{biretta91,marshall02} and Cen\,A
\citep{kraft02,hardcastle07}.

As for the BL\,Lac object Mrk\,421, the black hole mass of which
is considered to be $\sim 10^{6}M_{\sun}$ \citep{xie98}
or more, one anticipates a current of $I\ga 10^{17}~{\rm A}$.
Regarding the strongly beaming flow with a narrow viewing angle
in which $\delta\sim\Gamma_{j}$, we find the order of
$i_{0}\sim 10^{4}\delta~{\rm A}$ (in the regime, also note the
scaling $\nu_{c}\sim b(i_{0}/e)$ in equation \ref{eq:8}).
For $\delta\leq 100$, we accordingly have $N\ga 10^{11}$,
which is much larger than
$(N_{\lambda\sim d})_{\rm max}\sim 10^{8}$ (Fig.~\ref{fig:4}b).
The relation $N\gg (N_{\lambda\sim d})_{\rm max}$,
which is in contrast to $N\sim (N_{\lambda\sim d})_{\rm max}$
for the aforementioned FR\,I sources, can also be
found in other BL\,Lacs (Mrk\,501, for example).
It is thus ensured that the jets accompanying these compact objects
have many filaments smaller than $d$ (and thereby, smaller than $D$),
consistent with the assumption given in \citet{honda08}.

\section{DISCUSSION AND CONCLUSIONS}
\label{sec:5}

Another signature of the 'quantization' of magnetized current
filaments can be speculatively found in Fig.~\ref{fig:2}.
By invoking an approximate scaling $B\propto D^{-1/2}$,
we find that the column energy density of the magnetic field
$\epsilon_{b}$, estimated as $\sim B^{2}D/8\pi$,
is nearly constant:
\begin{equation}
\epsilon_{b}\left(={\rm const.}\right)
\sim 10^{11\sim 12}~{\rm ergs~cm^{-2}}.
\label{eq:14}
\end{equation}
\noindent
Obviously, this is at odds with the intuitive prediction of
$\epsilon_{b}\propto D^{-2}$ reflecting three-dimensional (3D)
adiabatic expansion, implying that 2D expansion, arguably in
the transverse plane, is preferentially inhibited due to
the pressure of the large-scale magnetic field that
would act radially inward \citep[e.g.][]{honda09}.
At the same time, it is also suggested that the inhomogeneous magnetic
fields in the interior of emission regions robustly sustain such an
averaged energy-density level against the self-contraction pressure.
Considering this aspect, the degenerate nature reflected in equation~(\ref{eq:14})
might be ascribed to the reasoning that any magnetized filaments
can by no means overlap spatially (or occupy the common space)
so as to carry a summed current exceeding the critical,
quantized value of $i_{0}$ [for $\Gamma_{j}\ga{\cal O}(1)$], to
prevent unlimited collapse \citep{honda00,honda00b}.
We note that this exclusive property could not be deduced from
the classical hydrodynamic description of current coalescence,
according to which an archetypical ${\bf J}\times{\bf B}$ pinch
leads to self-similar collapse \citep[e.g.][]{zhdanov98}.
The analogy to the updated notion can be found in the real mechanism
for the Fermi pressure in degenerate matter, based on the Pauli
exclusion principle \citep[e.g.][]{chandrasekhar67}.
With this insight, the $B-D$ property revealed in
Fig.~\ref{fig:2} can be regarded as a macroscopic appearance of
the quantization in relativistic flowing plasmas (Fig.~\ref{fig:1}),
and the beamed emission with $\nu_{\rm c}$ ($\la i_{0}/e$;
Section~\ref{sec:4}) as an acute appearance of collective quanta.
The origin of the $\epsilon_{b}$ value as such
(in equation~\ref{eq:14}) is another issue of note,
but the details are beyond the scope of this paper.

In conclusion, in the context of the filamentary jet model, we have
analytically provided a new scaling of the cut-off frequency $\nu_{c}$
of non-diffuse synchrotron emissions (equation~\ref{eq:10}).
The arguments expanded here are based on the plausible assumption
that small-amplitude magnetic fluctuations are superimposed on the
local mean fields that permeate through current filaments.
It has been demonstrated that the $\nu_{c}$ level
could be much higher than the previously suggested one
(equation~\ref{eq:1}), consistent with many X-ray measurements.
The expected range of the present break frequency (equation~\ref{eq:6})
appears to be comparable to the previous 'cut-off'
(in the nomenclature of earlier literatures).
In terms of the validity, we have provided extended discussions
of the smearing effect on the synchrotron spectrum at
$\nu_{c}$, the inverse Comptonization effect and the ageing
effect, which could involve the degenerative redshift of $\nu_{c}$.

By combining the results obtained with key observational data of
AGN jets, we have found the property that the minimum correlation
length of filamentary turbulence is larger for larger objects.
In particular, we address the issue that the correlation data for compact
BL~Lacs and FR\,I radio sources are likely aligned in a common
scaling, probably reflecting the spatiotemporal evolution of AGNs.
This trend is amenable to the AGN unification scheme, in which
the FR\,I sources are envisaged as misaligned BL~Lacs.
From translating the correlation scaling to the number scaling of filaments,
we could infer how the filaments are packaged in the jet interior,
and extract the consequence that the current carrying AGN jets
must possess many filaments smaller than the entire emission-region
size; e.g. for a pc-scale blazar jet the
number of filaments is $\sim 10^{11}$ and more.
The findings, including a constant column magnetic
energy density, support the conjecture that
AGN jets ubiquitously involve filamentary morphology.\\

\bsp

\label{lastpage}

\end{document}